# Observation of Weyl point pair annihilation in a gyromagnetic photonic crystal

Gui-Geng Liu[1,#], Zhen Gao[2,#], Peiheng Zhou[3,*], Qiang Wang[1], Yuan-Hang Hu[3], Maoren Wang[3], Chengqi Liu[3], Xiao Lin[4], Shengyuan A. Yang[5], Yihao Yang[4,*], Yidong Chong[1,6,*], and Baile Zhang[1,6,*]

[1]Division of Physics and Applied Physics, School of Physical and Mathematical Sciences, Nanyang Technological University, 21 Nanyang Link, Singapore 637371, Singapore.

[2]Department of Electrical and Electronic Engineering, South University of Science and Technology, Shenzhen 518055, China.

[3]National Engineering Research Center of Electromagnetic Radiation Control Materials, Key Laboratory of Multi-spectral Absorbing Materials and Structures of Ministry of Education, University of Electronic Science and Technology of China, Chengdu 610054, China.

[4]Interdisciplinary Center for Quantum Information, State Key Laboratory of Modern Optical Instrumentation, ZJU-Hangzhou Global Science and Technology Innovation Center, Zhejiang University, Hangzhou 310027, China.

[5]Research Laboratory for Quantum Materials, Singapore University of Technology and Design, Singapore 487372, Singapore.

[6]Centre for Disruptive Photonic Technologies, The Photonics Institute, Nanyang Technological University, 50 Nanyang Avenue, Singapore 639798, Singapore.

[#]These authors contributed equally to this work.

[*]E-mail: *phzhou@uestc.edu.cn (P.Z.); yangyihao@zju.edu.cn (Y.Y.); yidong@ntu.edu.sg (Y.C.); blzhang@ntu.edu.sg (B.Z.)*

**Weyl semimetals are gapless three-dimensional (3D) phases whose bandstructures contain Weyl point (WP) degeneracies[1]. WPs carry topological charge and can only be eliminated by mutual annihilation, a process that generates the various topologically distinct 3D insulators[1,2]. Time reversal (T) symmetric Weyl phases, containing a minimum of four WPs, have been extensively studied in real materials[3,4], photonic metamaterials[5,6], and other systems. Weyl phases with a single WP pair – the simplest configuration of WPs – are more elusive as they require T-breaking[7]. Here, we implement a microwave-scale gyromagnetic 3D**

**photonic crystal, and use field-mapping experiments to track a single pair of ideal WPs whose momentum space locations depend strongly on the biasing magnetic field. By continuously varying the field strength, we observe the annihilation of the WPs, and the formation of a 3D Chern insulator[2,8,9], a previously unrealised member of the family of 3D topological insulators (TIs). Surface measurements show, in unprecedented detail, how the Fermi arc states connecting the WPs evolve into TI surface states.**

One of the most remarkable discoveries in the physics of materials over the past two decades is that materials can be usefully characterised by the topological features of their energy bands[10,11]. We now know of a variety of TI phases, distinct from normal insulators and generated from them through band inversions (the closing and reopening of band gaps)[10,11]. The way band inversions happen is dimension dependent: in one- and two-dimensional materials, they occur just by tuning through a critical point, whereas in 3D they are generally mediated by Weyl phases and other topological semimetal phases[1,2]. The various topologically distinct 3D insulators can be regarded as being generated through the annihilation of WPs in various combinations, subject to various symmetry restrictions. With T unbroken, the annihilation of four WPs can produce a normal insulator, a weak TI, or a strong TI; with T broken, the annihilation of two WPs can produce a normal insulator or a 3D Chern insulator[2]. Despite its conceptual importance, WP annihilation has been challenging to observe experimentally. For instance, there have been experiments on using pressure to tune between a T-symmetric Weyl phase and a normal insulator[12,13], with the two phases distinguished by their transport characteristics, but the WPs and their topological surface states could not be resolved. No such WP tuning experiments have been performed on existing T-broken Weyl phases[7,14,15,16,17] to date.

Classical wave metamaterials such as photonic crystals[18] have proven to be excellent platforms for studying TIs, Weyl phases, and other topological phases. Compared to real materials, they offer considerable advantages in ease of fabrication and experimental characterisation. The emerging field of topological photonics[18] also has promising applications in topological lasers[19,20,21], optical delay lines[22], robust on-chip communications[23], and more. Notably, in microwave photonics, gyromagnetic materials offer a way to access T-broken phases[24], including T-broken Weyl phases that can host a single pair of WPs[25]. Recently, a Weyl phase with a single pair of WPs has been achieved in a lattice of cold atoms[7], but no WP annihilation was observed in that work.

Here, we report on the experimental implementation of a single pair of photonic Weyl points and their annihilation to form a 3D Chern insulator, based on a gyromagnetic photonic crystal operating at microwave frequencies. The photonic bandstructure can be tuned via the strength of an external static magnetic field, which can be continuously varied. At low field strengths, we show that the bandstructure hosts a single pair of WPs that are "ideal", meaning that there are no other bands over a significant range of nearby frequencies (a set of four ideal WPs has previously been observed in a non-magnetic photonic crystal[6], but the minimum set of two ideal WPs has only previously been achieved in a cold atom system[7]). As the field strength is increased, we track the WPs in momentum space through field mapping and show that the Weyl points cross the Brillouin zone (BZ) to annihilate, opening a complete bandgap. The resulting gapped phase is a 3D Chern insulator[2,8,9], which has never previously been observed in any physical setting. Using field measurements on the sample surface, we resolve the Fermi arc surface states associated with the WPs and show that they evolve, after WP annihilation, into robust surface states that circulate around the edges of the stack; this is another phenomenon that has been predicted theoretically but never before observed directly in experiment[2].

As depicted in Fig. 1a, the photonic crystal has a unit cell that consists of a gyromagnetic rod and a metallic plate perforated with holes (see Methods for details). The holes, positioned at the two sets of inequivalent corners of the hexagonal unit cell, introduce interlayer coupling and break the in-plane inversion symmetry. By applying a uniform external magnetic field along the $z$-axis, the underlying T symmetry of the gyromagnetic photonic crystal is broken. The strength of the biasing magnetic field serves a key degree of freedom for tuning the photonic bandstructure.

With relatively weak magnetic fields (such as 0.2 Tesla), the photonic crystal exhibits a magnetic Weyl phase featuring a single pair of WPs in the 3D BZ (left panel of Fig. 1b). As shown by the calculated band diagram in Fig. 1c, two bands linearly intersect at momenta ($k_x$, $k_y$, $k_z$) = ($4\pi/3a$, 0, $\pm 0.53\pi/h$), forming a single pair of ideal WPs that are related by the horizontal mirror symmetry. The existence of a single pair of WPs, the minimum possible, can only occur in a T broken system[1,7,25]. Since the WPs carry opposite topological charges, their projections on the surface BZ are connected by a Fermi arc of surface states, as shown in Fig. 1b. Simulations show that as the magnetic field strength is gradually increased, the system remains in the Weyl phase and the two WPs move in opposite directions along the *KH* high-symmetry line (as indicated by the two arrows in Fig. 1b), and the Fermi arc lengthens. The horizontal mirror symmetry guarantees that the two WPs must move in a symmetry way and finally can merge with each other. When the magnetic field strength reaches the critical value of 0.355 Tesla, the WPs merge at the high-symmetry point *H*, and the Fermi arc stretches all the way across the surface BZ (in Fig. 1b, it is depicted at 0.35 Tesla, just before the critical point). With a further increase of the magnetic field strength, the WPs disappear. This corresponds to a topological transition into a nontrivial insulating phase—a 3D Chern insulator. Figure 1d shows the calculated band diagram at 0.45 Tesla bias, which hosts a 3D topological bandgap from 19.3 to 19.7 GHz. The surface states once associated with the Fermi arc remain intact; now spanning the surface BZ, they are identified as the topological surface states of the 3D Chern insulator.

The topological properties of both the WPs and the 3D Chern insulator bands can be characterised by Chern numbers. The WPs have opposite Chern numbers (topological charges) of ±1, as indicated in Fig. 1b. Each band of the 3D Chern insulator is labeled by a triad of Chern numbers $C=(C^x, C^y, C^z)$, with each Chern number defined on the $\hat{x}$, $\hat{y}$, and $\hat{z}$ momentum planes[18,26,27] (see Methods). Our calculations show that $C=$ (0, 0, 1) for the bands below the gap, verifying that this is a 3D Chern insulator phase. The relationship between the topological charges of the WPs and the topological invariants of the 3D Chern insulator can be understood using a toy model composed of stacked Haldane models[28] (see Methods).

The experimental sample is shown in Figs. 2a and b. It is magnetically biased along the *z* axis by an external electromagnet (see Methods). To measure bulk transmission within the sample, we insert two dipole antennas (acting as a probe and a source, respectively) into the bulk, using the experimental setup shown in Fig. 2c. The point-like source excites almost all wavevectors in the bulk. By continuously tuning the biasing magnetic field strength from 0.175 to 0.475 Tesla, we obtain the bulk transmission plot of Fig. 2d, which clearly shows the opening of a complete photonic bandgap above 0.36 Tesla. This is the predicted topological transition from a gapless Weyl phase to an insulating phase. For comparison, the first-principles numerical calculation shows a gapless Weyl phase below 0.355 Tesla (with the Weyl frequencies indicated by the solid white line in Fig. 2d) and a 3D Chern insulator phase above 0.355 Tesla (with a complete 3D bandgap bounded by the two dashed white lines in Fig. 2d). The experimental and numerical results are in good agreement.

To reveal the topological properties of the two phases, we probe for topological surface states on each side of the transition. We cover the front (010) surface (parallel to the *xz* plane) of the crystal with a copper cladding, which acts as a trivial bandgap material, while wrapping the other surfaces with microwave absorbers. A point source is placed near the center of the front (010) surface to excite surface states. The field distributions on the surface are mapped out using a near-field scanning probe (see Methods). After applying the Fourier transform to the measured field distributions, we obtain the surface dispersion contours as a function of frequency.

With a 0.2 Tesla biasing magnetic field, for which the crystal is in the Weyl phase (Fig. 2d), the measured isofrequency contour at the Weyl frequency of 19.0 GHz is shown in Fig. 3a. We see a single open arc connecting the projections of the two WPs, at $(k_x, k_z) = (4\pi/3a, \pm 0.53\pi/h)$, consistent with the predictions of Fig. 1b,c. As we increase the magnetic field strength, the WPs move towards the boundary of the surface BZ, as shown in Fig. 3b (for 0.3 Tesla) and 3c (for 0.35 Tesla). At 0.45 Tesla, the crystal has entered the 3D Chern insulator phase, and Fig. 3d shows the measured surface isofrequency contour at 19.6 GHz (within the newly-opened bandgap), which consists of a curve that wraps around the surface BZ.

To demonstrate that the annihilation of the WPs is responsible for the opening of the topological complete bandgap, Figs. 3e and 3g show the frequent-dependent surface dispersion curves within different $k_z$ slices for the two topological phases. In the Weyl phase (at 0.2 Tesla), varying $k_z$ causes a topological transition (band inversion) of the projected 2D bands; the transition point corresponding to the WP, and the topological surface states exist on one side, from $k_z = 0$ to $k_z = 0.53\pi/h$, and are absent from $k_z = 0.53\pi/h$ to $k_z = 1\pi/h$. By contrast, in the 3D Chern insulator phase (at 0.45 Tesla), the projected bands are gapped for all $k_z$, with topological surface states spanning the gap. Moreover, the surface states possess positive group velocities along the $k_x$ direction, indicating chiral propagation (the plots for negative $k_z$ are similar and are thus omitted). The numerically calculated surface band structures, shown in Figs. 3f and 3h, are consistent with the experimental observations.

Next, we perform surface transmission experiments to probe the chiral propagation characteristics and robustness of the 3D Chern insulator's surface states. We cover the front (010), left (100) and right (100) surfaces of the sample with copper cladding, and all other surfaces with microwave absorbers. A point source is placed near the left (100) surface. With the sample in the 3D Chern insulator phase, Fig. 4a shows the experimentally-obtained field intensity distribution, revealing that the topological surface states propagate across the two sharp edges without significant reflection. When we further insert metallic obstacles along the path of the surface waves (see Methods), as shown in Fig. 4b, the surface waves bypass the obstacles with negligible reflection (note that in the Weyl phase, the surface states would be partially scattered into the gapless bulk[29,30]).

Finally, we study the refraction of the topological surface states at a termination of the photonic crystal into the surrounding environment. To facilitate measurement, we connect a parallel-plate waveguide to the front (010) surface (see Methods). As shown in Fig. 4c, a point source placed near the front (010) surface forms an image in the empty waveguide, through negative refraction at the termination (see phase matching analysis in Fig. 4d), with no observable reflection. Similar topological negative refraction phenomena have previously been observed in other settings[29,30], but without the direct imaging shown in Fig. 4c.

These results constitute clear experimental evidence of the annihilation of a single pair of Weyl points. The evolution and mutual annihilation of WPs, though pivotal to our understanding of 3D topological phases and how they are related to each other, had primarily been regarded as an abstract theoretical concept; our ability to perform a detailed experimental study of these processes is attributable to the gyromagnetic photonic crystal platform. The result of the Weyl point annihilation is the formation of a 3D Chern insulator, a 3D TI phase that has hitherto never been realised in any experimental setting. In the future, it would be interesting to improve our control over Weyl points in photonic crystals, such that we may be able to induce different Weyl point annihilations, generating topologically distinct insulator phases, on demand.

## Methods

**Materials and experimental setup.** The gyromagnetic material is yttrium iron garnet (YIG) ferrite with relative permittivity 14.3 and dielectric loss tangent 0.0002. The measured saturation magnetization is $M_s$ = 1780 Gauss, and the gyromagnetic resonance loss width is 35 Oe. The copper plates are prepared by coating copper with thickness of 0.035 mm onto Teflon woven glass fabric laminate. The dielectric foam (ROHACELL 31 HF) used to fix the position of the YIG rods has relative permittivity 1.04 and loss tangent 0.0025. The air holes penetrate through the copper plate layers and the dielectric foam layers, as shown in Supplementary Fig. 1a. In the measurements, uniform external magnetic fields generated by an electromagnet are applied along the $z$ axis, producing a strong gyromagnetic response in the ferrite rods, as shown in Supplementary Fig. 1b. The spatial non-uniformity of the magnetic fields is less than 2% in the sample region, measured and calibrated using a gaussmeter (HFBTE, TD8620). To produce the results shown in Fig. 4b, several copper pillars are inserted into the coupling holes to function as the metallic obstacles, as shown in Supplementary Fig. 1c. As shown in Supplementary Fig. 1d, a parallel-plate waveguide is installed near the front (010) to produce the results shown in Fig. 4c. During the measurements, two microwave antennas with diameters 1.2 mm are connected to a vector network analyzer (Agilent Technologies, E8363C) to function as the source and the probe, respectively. Complex field distributions are measured by inserting the probe into the air holes one-by-one and scanning along $z$ direction in small steps.

**Numerical simulation.** The bulk and surface band diagrams are numerically simulated with a finite element method solver (COMSOL Multiphysics RF Module). The copper plates are modelled as perfect electric conductors. The surface dispersions are calculated in a supercell with 40 cells in the $y$ direction, where two (010) boundaries are set as perfect electric conductors and the other boundaries have periodic boundary conditions.

The relative permeability tensor of the gyromagnetic materials has the form $\tilde{\mu} = \begin{bmatrix} \mu_r & i\kappa & 0 \\ -i\kappa & \mu_r & 0 \\ 0 & 0 & 1 \end{bmatrix}$, where $\mu_r = 1 + \frac{(\omega_0 + i\alpha\omega)\omega_m}{(\omega_0 + i\alpha\omega)^2 - \omega^2}$, $\kappa = \frac{\omega\omega_m}{(\omega_0 + i\alpha\omega)^2 - \omega^2}$, $\omega_m = \gamma\mu_0 M_s$, $\omega_0 = \gamma\mu_0 H_0$, and $\mu_0 H_0$ is the external magnetic field along the $z$ direction, $\gamma = 1.76 \times 10^{11}$ s$^{-1}$T$^{-1}$ is the gyromagnetic ratio, $\alpha$ = 0.0088 is the damping coefficient, and $\omega$ is the operating frequency. The dispersion of the permeability tensor elements $\mu_r$ and $\kappa$ for 0.2 Tesla and 0.45 Tesla biased magnetic fields are shown in Supplementary Fig. 2a and 2b, respectively. Since the resonance frequency of the permeability is far from the Weyl frequency or the bandgap, we neglect the weak dispersion in the frequency range of interest and assume a dispersionless permeability to facilitate simulation. Specifically, the dispersionless $\mu_r$ and $\kappa$ are taken from the values of the dispersive $\mu_r$ and $\kappa$ at the Weyl frequency in the gapless phase, or at the mid-gap frequency in the gapped phase. For 0.2 Tesla bias, the Weyl frequency is 19.0 GHz and we obtain $\mu_r$ = 0.92 and $\kappa$ = -0.29. For 0.45 Tesla bias, the mid-gap frequency is 19.5 GHz and we obtain $\mu_r$ = 0.72 and $\kappa$ = -0.44. The values of $\mu_r$ and $\kappa$ at other magnetic field strengths are shown in Supplementary Fig. 2c.

**Topological characterization of the gyromagnetic photonic crystal.** The topological charge of a Weyl point is calculated by integrating the Berry curvatures over a surface enclosing the Weyl point. The Berry curvature is defined as $\Omega_n(\mathbf{k}) = \nabla_\mathbf{k} \times A_n(\mathbf{k})$ , where $A_n(\mathbf{k}) = i\langle u_{n,\mathbf{k}} | \nabla_\mathbf{k} | u_{n,\mathbf{k}} \rangle$ is the Berry connection and $u_{n,\mathbf{k}}$ is the Bloch state for the $n$-th band. The topological charge of a Weyl point is given by $\frac{1}{2\pi}\int_S \Omega_n(\mathbf{k}) d\mathbf{s}$, where $S$ is a surface enclosing the Weyl point and $n$ = 2 for our case. For photonic crystal with 0.2 Tesla bias, the calculated topological charges of the Weyl points are indicated in Fig. 1b and Supplementary Fig. 3a. In addition, a Chern number is well defined for a 2D slice of the first BZ not containing the Weyl point. Specifically, the Chern numbers for the $\hat{x}$, $\hat{y}$, and $\hat{z}$ momentum planes are defined as $C_n^{x,y,z} = \frac{1}{2\pi}\int_{S^{x,y,z}} \Omega_n^{x,y,z} d^2k$ , where $S^{x,y,z}$ are the $k_{x,y,z}$-fixed 2D BZ. For example, the calculated $C^z$ for the second band of the photonic crystal with 0.2 Tesla biased is shown in Supplementary Fig. 3b, showing that $C^z(|k_z| < 0.53\pi/h) = 1$ and $C^z(|k_z| > 0.53\pi/h) = 0$. For the 3D Chern insulator, the system is globally gapped, so the Chern numbers $C_n^{x,y,z}$ are well defined over the entire BZ. The calculated Chern numbers for the second band of the photonic crystal with 0.45 Tesla bias are $C^z = 1$ and $C^{x,y} = 0$, as shown in Supplementary Fig. 3c. The first band is always trivial. Hence, the complete bandgap between the second and the third bands of the 3D photonic Chern insulator can be characterised by a triad of Chern numbers $C=(C^x, C^y, C^z) = (0,0,1)$.

**Three-dimensional Stacked Haldane model.** A 3D stacked Haldane model is constructed by AA-stacking the 2D Haldane model[28] and introducing the interlayer coupling (lattice constant along $z$-axis is $h$), as demonstrated in Supplementary Fig. 4a. For each layer, the A (purple) and B (orange) sublattice sites possess on-site energies of +M and –M, respectively. There is a real hopping term $t_1$ between the nearest neighbour sites (solid black line) and a complex hopping term $t_2\exp(i\phi)$ between the next nearest neighbour sites (solid purple and orange lines). The interlayer hopping terms for A and B sublattice sites are $t_a$ (purple dashed line) and $t_b$ (orange dashed line), respectively. The Bloch Hamiltonian of the 3D Stacked Haldane model is given by $H(\mathbf{k})=\begin{pmatrix}\Delta^+ & \Phi^- \\ \Phi^+ & \Delta^-\end{pmatrix}$, where $\Delta^{\pm}=2t_2\sum_{i=1,2,3}\cos(\phi\pm\mathbf{k}\cdot\mathbf{b}_i)\pm M+2t_a\cos(k_z h)$, $\Phi^{\pm}=t_1\sum_{i=1,2,3}e^{\pm i\mathbf{k}\cdot\mathbf{a}_i}$, ($\mathbf{a}_1$, $\mathbf{a}_2$, $\mathbf{a}_3$) are the displacements from a B site to its three adjacent A sites, and $\mathbf{b}_1=\mathbf{a}_2-\mathbf{a}_3$, $\mathbf{b}_2=\mathbf{a}_3-\mathbf{a}_1$, and $\mathbf{b}_3=\mathbf{a}_1-\mathbf{a}_2$. By tuning the hopping terms, this 3D model exhibits several phases, as shown in Supplementary Fig. 4b: a 3D Chern insulator, a 3D trivial insulator, and gapless Weyl phases. The brown and grey regions indicate the Weyl phases, which feature either four or two Weyl points in the first BZ, respectively.

## Data availability

The data that support the plots within this paper and other findings of this study are available from the corresponding author upon reasonable request.

## Acknowledgments

This work was sponsored by the Singapore Ministry of Education Academic Research Fund Tier 3

Grant MOE2016-T3-1-006, and the National Research Foundation Competitive Research Program NRF-CRP23-2019-0007. Work at University of Electronic Science and Technology of China was sponsored by the National Natural Science Foundation of China (NSFC) (Nos. 52022018 and 52021001).

## Authors Contributions

Y.Y. and G.-G.L. initiated the idea. Q.W. preformed the tight-binding calculation. G.-G.L. performed the simulation. G.-G.L., Z.G. and Y.Y. designed the experiments. Z.G. and P.Z. fabricated samples. P.Z., Y.-Y.H., M.W and C.L. carried out the measurements. G.-G.L. analysed data. G.-G.L., Y.Y., P.Z., Z.G., Q.W., X.L., S.A.Y., Y.C. and B.Z. wrote the manuscript. B.Z., Y.C., Y.Y. and P.Z. supervised the project.

## Competing Interests

The authors declare no competing interests.

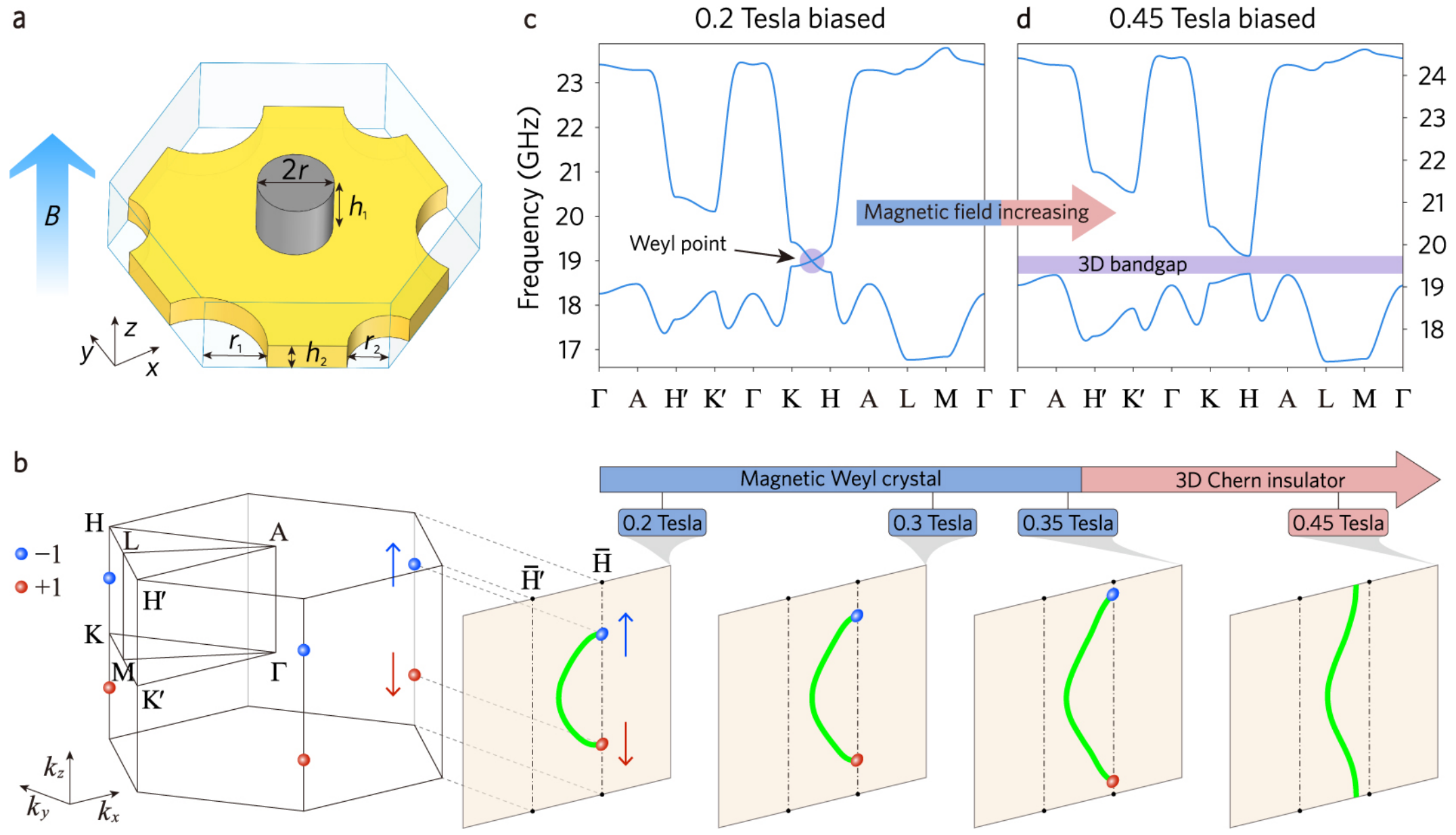

**Fig. 1 | Design of a single pair of Weyl points and their annihilation procedure in a gyromagnetic photonic crystal. a**, Unit cell of the gyromagnetic photonic crystal biased by an external magnetic field *B*. The lattice constant in the *xy*-plane is $a$ = 10 mm. The periodicity along $z$ axis is $h$ = 3 mm. The gyromagnetic cylinder (colored in grey) has radius $r$ = 1.2 mm, and height $h_1$ = 2 mm. The copper plate (colored in yellow) separating the gyromagnetic cylinder has thickness $h_2$ = 1 mm. The coupling holes at the two sets of inequivalent corners have radius $r_1$ = 2 mm and $r_2$ = 1.3 mm, respectively. **b**, The bulk Brillouin zone and (010) surface Brillouin zone. The blue and red dots are the two ideal Weyl points with opposite topological charges, when the biased magnetic fields are relatively weak. The red and blue arrows indicate the moving directions of the Weyl points with the enhancement of biased magnetic fields. The green curves on the surface BZ indicates a photonic Fermi arc for 0.2 Tesla, 0.3 Tesla, 0.35 Tesla, and 0.45 Tesla biased magnetic fields, respectively. **c**-**d**, Band diagrams of the photonic crystal under biased magnetic fields of 0.2 and 0.45 Tesla, respectively. The purple dot in **c** denotes a Weyl point. The purple rectangle in **d** represents a complete 3D topological bandgap from 19.3 to 19.7 GHz.

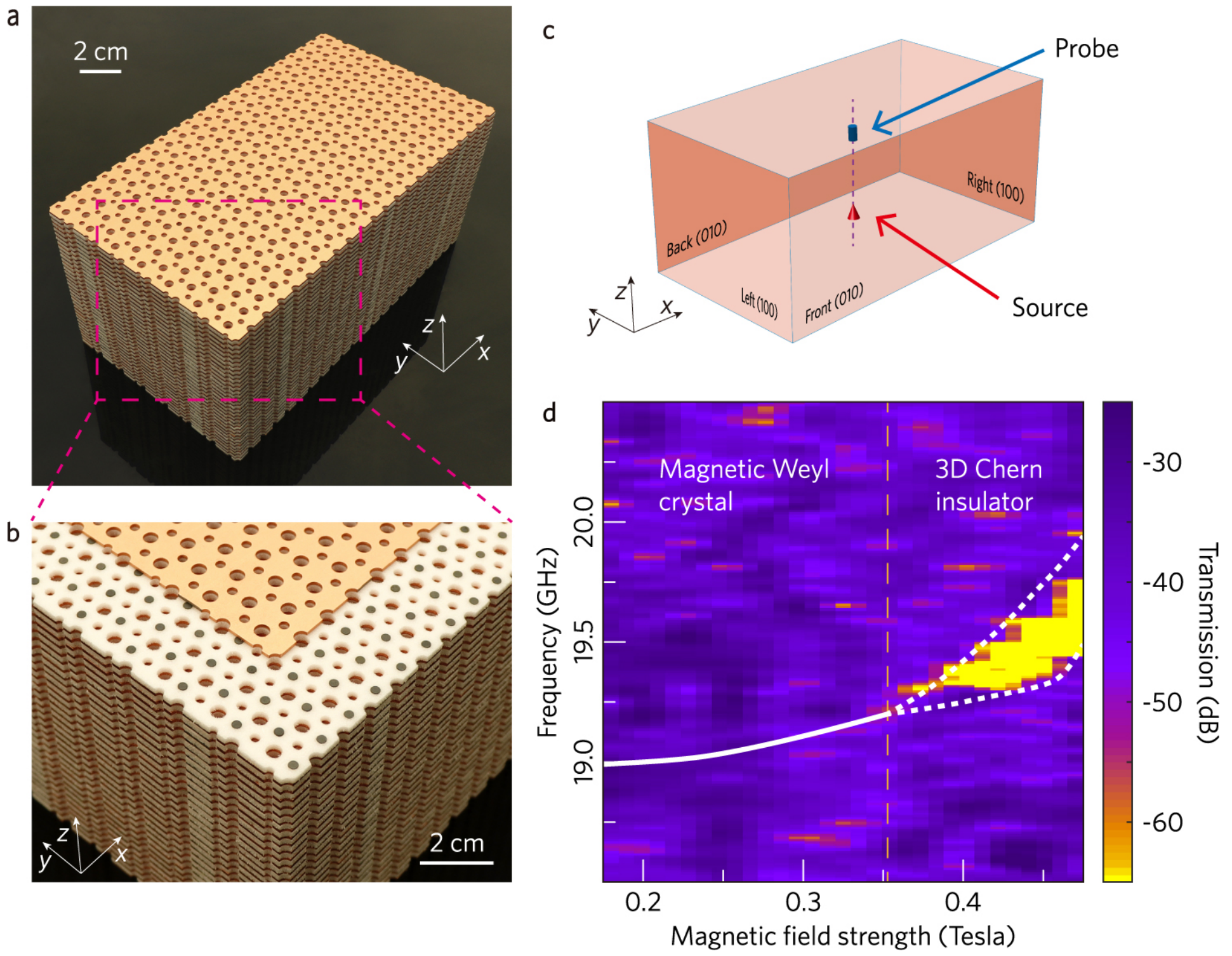

**Fig. 2 | Observation of bandgap opening in the gyromagnetic photonic crystal. a**, Photograph of the fabricated experimental sample. The sample has 33 layers in the $z$ direction and 22 (14) unit cells in the $x$ ($y$) direction. **b**, Zoomed view of the sample, where the first copper plate on the top has been shifted for visualization. The gyromagnetic rods are embedded into dielectric foams to fix their positions. **c**, Experimental setup for measurements of the bulk transmission. The source and the probe are inserted from the center of the bottom and top surfaces, with 2 cm depth. All surfaces are wrapped with microwave absorbers during measurements. **d**, Measured bulk transmission under different biased magnetic fields. The white solid line indicates the calculated frequency of the Weyl points. The two white dashed lines enclose the complete 3D bandgap. The phase transition from a gapless Weyl crystal to a gapped 3D Chern insulator occurs at the magnetic field of 0.355 Tesla.

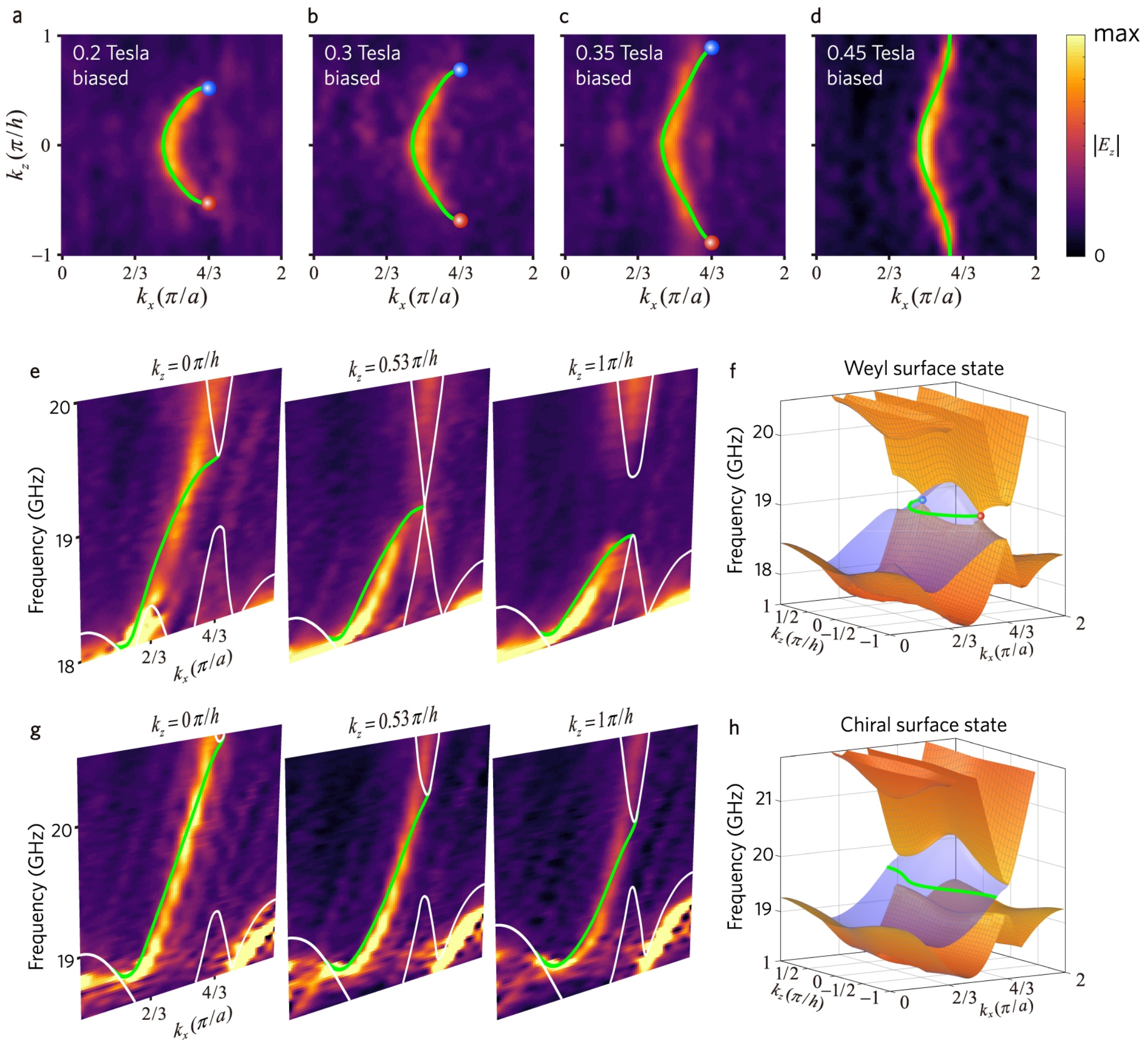

**Fig. 3 | Observation of topological surface states before and after the Weyl point annihilation. a-d**, Measured surface dispersion contours on the front (010) surface under biased magnetic fields of 0.2 Tesla at 19.0 GHz in **a**, 0.3 Tesla at 19.1 GHz in **b**, 0.35 Tesla at 19.15 GHz in **c**, and 0.45 Tesla at 19.6 GHz in **d**, respectively. The green line presents the calculated Fermi arc. The red and blue dots correspond to the projected Weyl points with opposite topological charges. **e,g**, Measured surface dispersions on the front (010) surface for fixed values of $k_z = 0$, $0.53\pi/h$, and $1\pi/h$, under biased magnetic field of 0.2 Tesla and 0.45 Tesla, respectively. **f,h**, Simulated band structure on the front (010) surface under magnetic field of 0.2 Tesla and 0.45 Tesla, respectively. The white and green curves in **e** and **g** indicate the simulated envelopes of the projected bulk dispersions and surface dispersions, respectively. The blue curved surfaces in **f** and **h** represent the topological surface states, while the orange sheets indicate the envelopes of the projected bulk dispersions. The green curves in **f** and **h** correspond to the same curves in **a** and **d**, respectively.

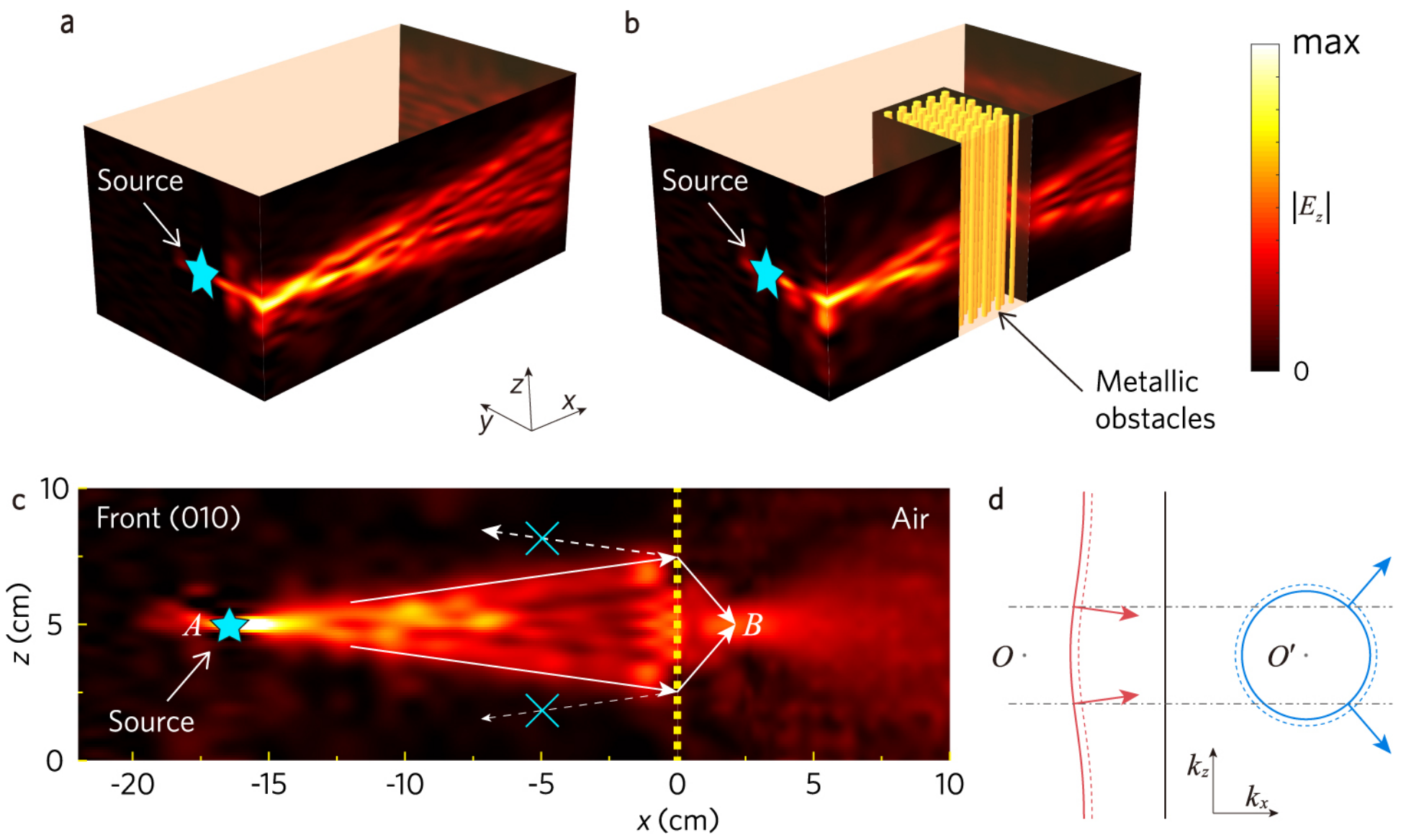

**Fig. 4 | Robustness of topological surface states in the 3D Chern insulator.** **a**, Measured field distribution of topological surface waves excited by a point source (cyan star) oscillating at 19.6 GHz. **b**, Measured field distribution in the same setup in **a**, except the copper pillars (yellow rods) inserted into the photonic crystal as metallic obstacles. **c**, Measured field distribution of topological surface waves excited by a point source placed near the front (010) surface and oscillating at 19.6 GHz. The yellow dashed line indicates the interface between the front (010) surface (left) and an additional parallel-plate waveguide (right), the latter of which serves as the surrounding environment. The surface waves from point *A* (the source) are focused at the image point *B* in the empty waveguide. **d**, Isofrequency contours of the topological surface states (left) and the fundamental mode in the empty waveguide (right). The solid and dashed lines represent dispersions at 19.6 GHz and a slightly higher frequency, respectively. The arrows indicate the propagating directions.

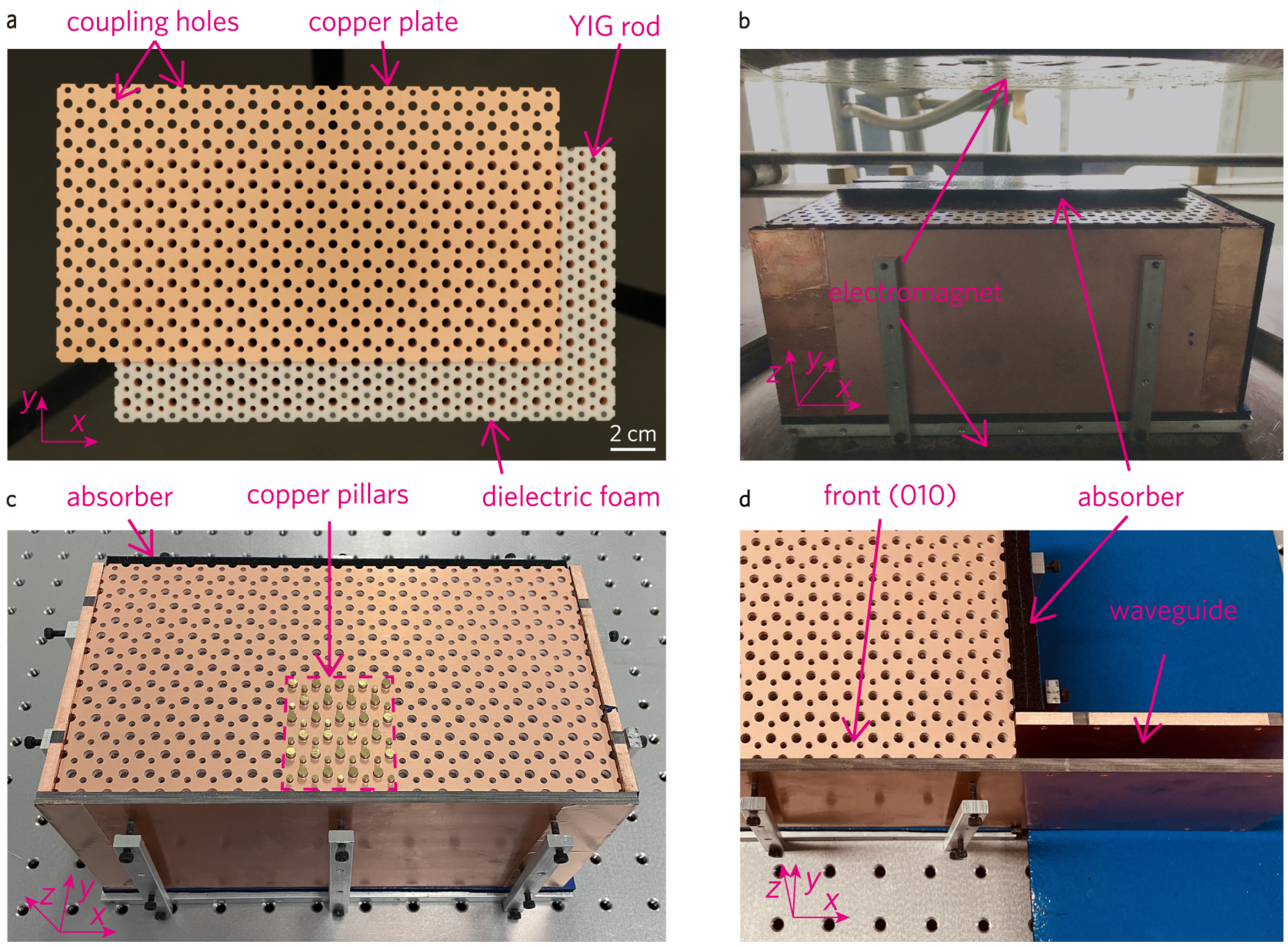

**Supplementary Fig. 1 | Experimental set-ups. a,** Top view of the fabricated sample, where the first copper plate on the top is shifted for visualization. **b,** Electromagnet used to produce magnetic fields. **c,** Copper pillars inserted into the coupling holes to function as metallic obstacles. **d,** Parallel-plate waveguide connected to the front (010) surface. The separation of the waveguide is 15 mm.

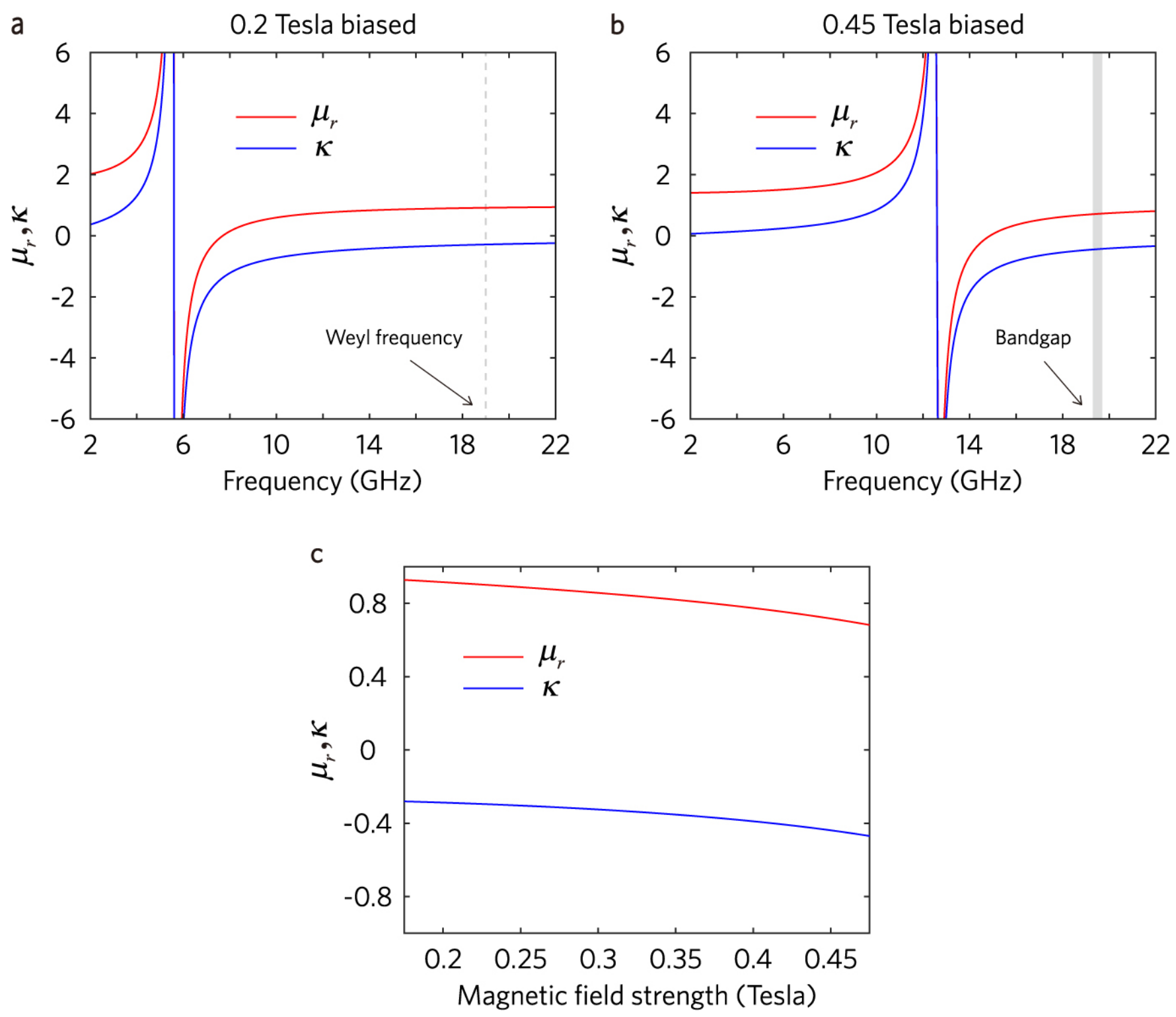

**Supplementary Fig. 2 | Permeability of the gyromagnetic materials. a-b,** Frequency-dependent elements $\mu_r$ and $\kappa$ of the permeability tensor of the gyromagnetic materials with the 0.2 Tesla and 0.45 Tesla biased magnetic fields, respectively. Grey dashed line in **a** indicates the Weyl frequency and grey rectangle in **b** indicates the bandgap of the photonic crystals. **c,** Dispersionless $\mu_r$ and $\kappa$ adopted in simulation as a function of biased magnetic fields.

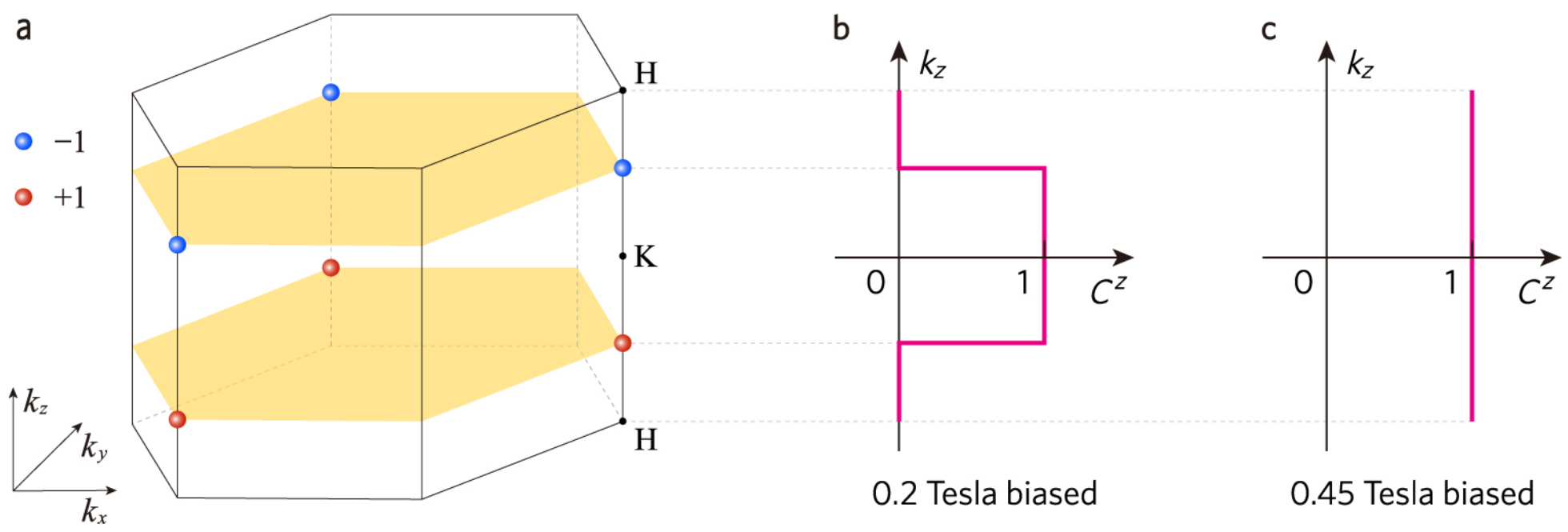

**Supplementary Fig. 3 | Chern number of the photonic crystal. a,** First bulk Brillouin zone of the photonic crystal. The blue and red dots are the two ideal Weyl points with opposite topological charges, when the biased magnetic field is 0.2 Tesla. **b-c,** Chern number of the 2D Brillouin zone with a fixed $k_z$ for the photonic crystal under biased magnetic fields of 0.2 Tesla and 0.45 Tesla, respectively.

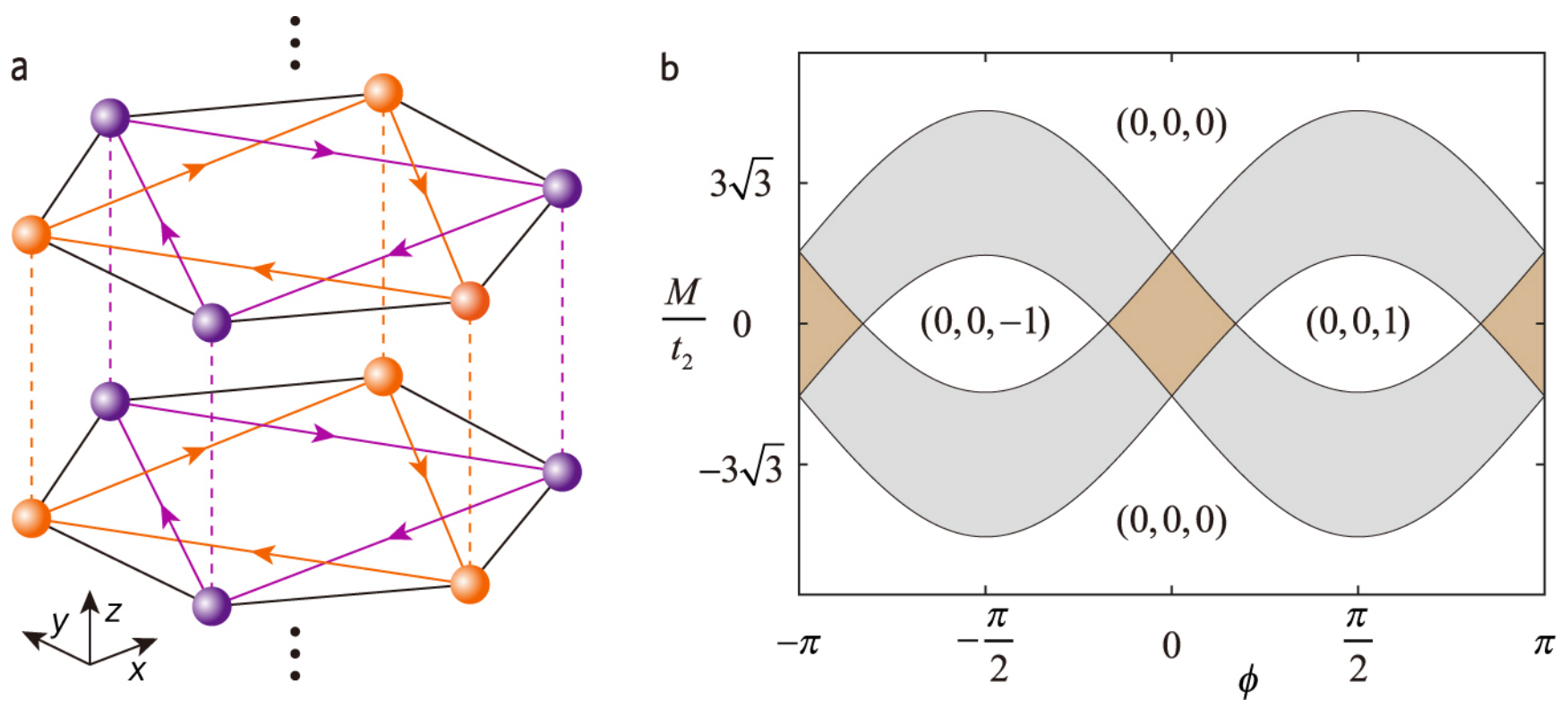

**Supplementary Fig. 4 | Three-dimensional stacked Haldane model. a,** 3D stacked Haldane model constructed by AA-stacking of a 2D Haldane model. The on-site energies are +M for A (purple spheres) and –M for B (orange spheres) sublattice sites, respectively. The black solid lines represent nearest hopping $t_1$, and the directional purple and orange solid lines represent next nearest hopping $t_2\exp(i\phi)$. The purple and orange dashed lines represent interlayer hopping $t_a$ and $t_b$, respectively. **b,** Phase diagram of the 3D stacked Haldane model. The gapped topological phases have been indicated by a triad of Chern insulator. The brown and the grey regions represent the gapless Weyl semimetal phases hosting four and two Weyl points in the first Brillouin zone, respectively. In the diagram, $t_2 = 1.2$, $t_a = 2$, and $t_b = -1.2$.